      [1999/12/01 v1.4c Il Nuovo Cimento]
\documentclass{cimento}
\usepackage{graphicx}
\usepackage{array}

\title{The double slit experiment and the time reversed \\fire alarm\thanks{Accepted March 2010 for publication in Il Nuovo Cimento B. The original publication will be available at http://www.sif.it/SIF/en/portal/journals}}
\author{T.~Halabi\thanks{halabit@radonc.ucsf.edu}}

\instlist{\inst{} University of California San Francisco - San Francisco, CA 94143-1708, USA. 
                         }
                         
\PACSes{\PACSit{03.65.Ta}{Foundations of quantum mechanics; measurement theory} \PACSit{05.20.-y}{Classical statistical mechanics}}
\begin{document}

\maketitle

\begin{abstract}
When both slits of the double slit experiment are open, closing one paradoxically increases the detection rate at some points on the detection screen. Feynman famously warned that temptation to ``understand'' such a puzzling feature only draws us into blind alleys. Nevertheless, we gain insight into this feature by drawing an analogy between the double slit experiment and a time reversed fire alarm. Much as closing the slit increases probability of a future detection, ruling out fire drill scenarios, having heard the fire alarm, increases probability of a past fire (using Bayesian inference). Classically, Bayesian inference is associated with computing probabilities of past events. We therefore identify this feature of the double slit experiment with a time reversed thermodynamic arrow. We believe that much of the enigma of quantum mechanics is simply due to some variation of time's arrow. In further support of this, we employ a plausible formulation of the thermodynamic arrow to derive an uncertainty in classical mechanics that is reminiscent of quantum uncertainty.
\end{abstract}







\section{Introduction}
\label{mp}
Facile views on time's arrow in physics impede progress in foundations of quantum mechanics. Consider the key role played by Bell's theorem, and the EPR experiments that followed, in shifting physicists' views on fundamental physics. Bell's theorem pivots heavily on the assumption of a unidirectional causal arrow of time\footnote{This is a major theme of \cite{Price96} in which Price writes: ``the theorem requires the assumption that the values of hidden variables are statistically independent of future measurement settings.''}. Yet, most physicists claim that their views on fundamental physics are time symmetric. There is some insincerity in either this claim of a time symmetric view or in the conclusions drawn from the EPR episode. The perceived innocence of Bell's assumption of a unidirectional casual arrow of time is rooted in physicists' continued stigma of the grandfather paradox. But hasn't this stigma been removed by Wheeler and Feynman \cite{Wheeler49} and Schulman \cite{schul01,schul71} already? 

We first began to suspect a role for a variant of time's arrow in quantum mechanics when considering the following peculiarity:

\begin{quote}
Quantum theory's prediction of the grotesque superpositions of the cat paradox is as outlandish as the experiment it prescribes to empirically confirm their existence: interference measurement of macroscopic superposition.
\end{quote}

If these interference measurements can be seen as giving rise to the superpositions they are designed to confirm, then the absence of grotesque superpositions from our experience can be attributed to the impracticality of these measurements. This would require a reversed causal arrow of time: the absence (presence) of \emph{future} interference measurement prevents (allows) \emph{present} states that evolve into macroscopic superposition.

Are we not free to choose any initial state we desire? Is it then not contradictory to allow the future to influence our freely chosen initial state? We submit that we do have choice over the ``course grain'' or macrostate attributes of the initial state but not the finer microstate. 
 
In support of this, Schulman \cite{schultext} has shown that it is always possible to fine tune present quantum states of system and environment so that they evolve clear of grotesque superposition. These fine tuned ``special'' initial microstates are very rare relative to ones that do evolve into grotesque superposition. Schulman's argument is consequently against the fundamental postulate of statistical mechanics, that of assigning equal a priori probability to all microstates of a given macrostate. 
In the context of classical mechanics we already know, however, that the practice of assuming equal a priori probability for the microstates is \emph{apparently} valid when projecting into the future but not into the past \cite{schul75}. This same practice is used in Boltzman's statistical argument for the second law of thermodynamics but, as is well known, incorrectly leads to the conclusion of a higher entropy past when projecting backwards in time.  

Schulman is therefore inclined to attribute the restriction to ``special'' initial states to a final boundary condition, that the universe not be in macroscopic superposition in the distant future. Along with a similiar inital boundary condition in the distant past these are called the ``two time boundary condition''. 


The retreat to an ad hoc boundary condition is rooted in the simple observation that most of the dynamical laws that we are aware of, such as Newton's second law or Maxwell's equations, impose no restrictions on the initial state. In passing, however, we point out one exception to this in the literature: in the ``shutter'' example \cite{Wheeler49} of Wheeler-Feynman absorber theory the shutter's initial state is restricted by the theory's dynamics. 

Much of the above represents the basic encounters a physicist makes with the subject of time's arrow when seriously contemplating the measurement problem. The next two sections examine more complex illustrations of the connection between these two themes. They attempt to answer the following question: how much of the enigma of quantum mechanics is merely that of time's arrow or variations thereof? In section \ref{uncertainty} we obtain an unavoidable uncertainty in classical mechanics, reminiscent of quantum uncertainty, by exploring a plausible formulation of the thermodynamic arrow. In section \ref{ds} the enigmatic feature of the double slit experiment (viewed forward in time) is shown to be that of classical Bayesian inference of the past. Section \ref{uncertainty} is not required reading for section \ref{ds} and the reader may begin with the latter.

Caves et al. \cite{caves02} have reached similar conclusions on the prominence of Bayesianism in quantum mechanics although through a starkly different path than ours. We contrast their approach with ours in Appendix \ref{Caves}.

\section{Unavoidable uncertainty in time symmetric classical mechanics} 
\label{uncertainty}
Just as unavoidable uncertainty is associated with quantum mechanics, unavoidable uncertainty would also arise in classical mechanics if the thermodynamic arrow of time is generalized.

The most conventional formulation of a thermodynamic arrow is obtained by requiring a single initial boundary condition that the system be in a (low entropy) macrostate $\epsilon_0$ at a past time $t= 0$. Schulman \cite{schul05} employs a more time symmetric framework by also including a final boundary condition, that the system be in macrostate $\epsilon_T$ at time $t = T$. The thermodynamic arrow then results from the stipulation that the time, $t$, at which we currently find our universe satisfies $0<t \ll T$. The probability that the system is in a macrostate $\Delta_{\alpha}$ at such a time is then given by:
\begin{equation}
\rho_{\alpha} = \frac{\mu(\phi^{(t)}(\epsilon_0) \cap \Delta_{\alpha} \cap \phi^{(t-T)} (\epsilon_T))}{\mu(\phi^{(t)}(\epsilon_0) \cap \phi^{(t-T)}(\epsilon_T))}, \label{p0}
\end{equation}
where $\phi^{t}(\cdot)$ is a measure preserving evolution of all phase space points in $(\cdot)$, and $\mu(\cdot)$ represents the measure (phase space volume) of $(\cdot)$. 

Knowing that the magnitude of $t-T$ is large, we assume perfect mixing to obtain the following approximation: $\mu(A \cap \phi^{(t-T)}(B)) = \mu(A) \mu(B)$, where $A$ and $B$ are any macrostates. Equation \ref{p0} then becomes
\begin{equation}
\rho_{\alpha} = \frac{\mu(\phi^t(\epsilon_0) \cap \Delta_{\alpha})}{\mu(\epsilon_0)} \label{pc}.
\end{equation}
From this equation one sees how the future conditioning $\epsilon_T$ has no effect on the present probability distribution of macrostates. The future conditioning does, however, restrict the set of microstates \footnote{This is reminiscent of the EPR experiment in which the future's effect is concealed in hidden variables. Here it is concealed in microstate variables.}.

From this point, Schulman goes on to derive the causal arrow of time \cite{schul05}. His remarkable derivation will not be reproduced here, but the reader is strongly encouraged to consult the original reference \cite{schul05}. In passing, we note that his is ultimately a derivation of our profound inability to remember the future. Remembering the future requires a time reversed causal arrow since a present memory (effect) must precede the memorable event (cause).

\begin{figure} 
\center
\includegraphics[scale = .35]{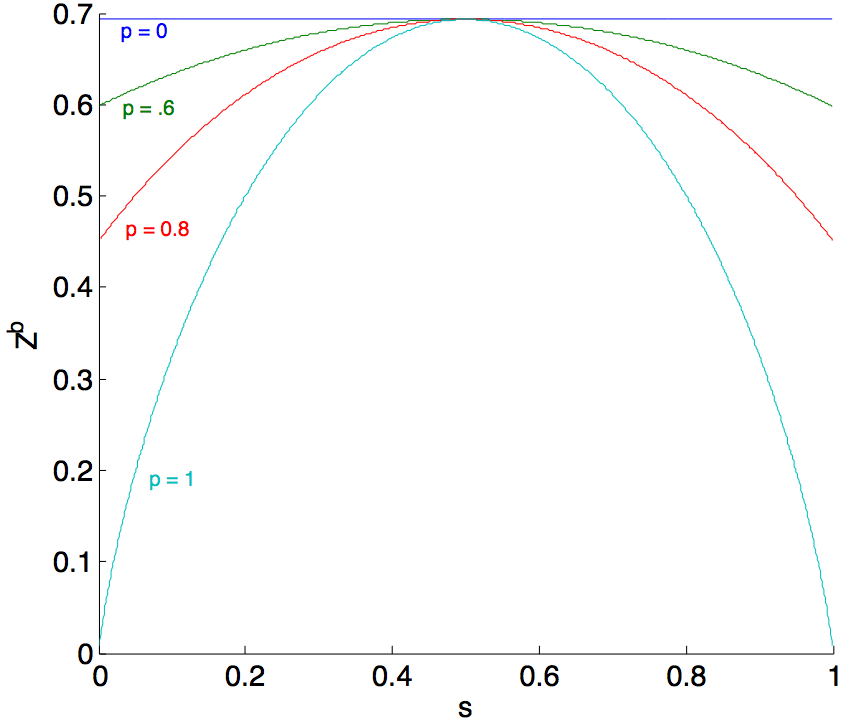}
\caption{Plot of the entropy under no mixing, $Z^b$, for the case $\alpha = 1,2$ (given by equation \ref{zb}). When $p = 1$ plot reduces to that of perfect mixing (equation \ref{pz}). }
\label{entPlot}
\end{figure}

Assume, for simplicity, that all macrostates have the same measure and choose units such that: $\mu(\Delta_{\alpha}) = 1,~ \forall \alpha$. To further simplify our analysis we assume the special case whereby 
\begin{equation}
\phi^t(\epsilon_0) = \phi^t(\epsilon_0) \cap (\Delta_1 \cup \Delta_2), \label{assume}
\end{equation}
i.e. $\Delta_1$ and $\Delta_2$ are exhaustive. From equations \ref{pc} and \ref{assume} we have: 
\begin{eqnarray}
\rho_1 &=& s \equiv \frac{\mu(\phi^t(\epsilon_0) \cap \Delta_1)}{\mu(\epsilon_0)},    \nonumber \\
\rho_2 &=& 1-s. \label{pm}
\end{eqnarray}
The entropy, given equations \ref{pm}, is:
\begin{equation}
Z = - s~ ln(s) -(1-s) ln(1-s). \label{pz}
\end{equation}
Its plot is shown in Figure \ref{entPlot} (case of $p = 1$).

Quantum experiments are marked by the high precision with which one seeks to determine the state of the system. Such circumstances motivate us to consider $\mu(\phi^t(\epsilon_0) \cap \Delta_{\alpha})$ so small as to have not undergone any mixing by time $T$. In other words, we now assume that $\phi^t(\epsilon_0) \cap \Delta_{\alpha}$ is either evolved whole into $\epsilon_T$ by $\phi^{T-t}$, in which case we have: 
\begin{equation}
\phi^t(\epsilon_0) \cap \Delta_{\alpha} \cap \phi^{t-T}(\epsilon_T) = \phi^t(\epsilon_0) \cap \Delta_{\alpha} \label{white}
\end{equation}
or it evolves whole outside of $\epsilon_T$, in which case we have:
 \begin{equation}
 \phi^t(\epsilon_0) \cap \Delta_{\alpha} \cap \phi^{t-T}(\epsilon_T) = \emptyset. \label{green}
 \end{equation}
 We regard the former as having a probability $p$, and the latter, $1-p$. Under the incomplete (in fact, none at all) mixing now assumed, we regard the probability of partial evolution of the very small $\phi^t(\epsilon_0) \cap \Delta_{\alpha}$ into $\epsilon_T$ as zero.

We will show that these simple assumptions of no mixing introduce bounds on the probabilities, prohibiting both $\rho_1$ and $\rho_2$ from reaching $1$ or $0$. We will use a superscript $b$ (``bound'' or ``blurred'') to indicate quantities for which these effects of incomplete mixing are being accounted for. For example, we will derive a lower bound on $Z^b$, which we regard as an unavoidable uncertainty (for our special case) in classical mechanics.

\begin{table}[tbp]
\caption{Summary of calculations for $\rho^b_1$ and $\rho^b_2$. Equation \ref{assume} was used in obtaining denominator of equation \ref{p0}.}
\centering
\begin{tabular}{ m{2.4cm}  | m{2.5cm} |  m{2.6cm} | m{2.6cm} c }
\hline \hline \\
 scenario                                                          & \centering $\Delta_1$ In, $\Delta_2$ In                                 & \centering $\Delta_1$ In, $\Delta_2$ Out                                     &  \centering  $\Delta_1$ Out, $\Delta_2$ In      &                \\ \hline 
 probability                                                       &    \centering  $\frac{p}{2-p}$                                                      & \centering $\frac{1-p}{2-p}$                                                             &  \centering  $\frac{1-p}{2-p}$             &                                      \\  \hline
 numerator of eq. \ref{p0} for $\Delta_1$     &    \centering  $\mu(\phi^t(\epsilon_0) \cap \Delta_1)$         & \centering $\mu(\phi^t(\epsilon_0) \cap \Delta_1)$                    &  \centering $0$                                                        &                  \\  \hline
 numerator of eq. \ref{p0} for $\Delta_2$     &   \centering   $\mu(\phi^t(\epsilon_0) \cap \Delta_2)$        & \centering $0$                                                                                    &   \centering $\mu(\phi^t(\epsilon_0) \cap \Delta_2)$     &       \\  \hline
 denominator of eq. \ref{p0}                           &    \centering  $\mu(\epsilon_0)$                                             & \centering $\mu(\phi^t(\epsilon_0) \cap \Delta_1)$                     & \centering   $\mu(\phi^t(\epsilon_0) \cap \Delta_2)$   &        \\  \hline
 contribution to $\rho^b_1$                                &   \centering   $\frac{p}{2-p}s$                                               & \centering $\frac{1-p}{2-p}$                                                             & \centering $0$                                                             &             \\  \hline
 contribution to $\rho^b_2$                                &    \centering  $\frac{p}{2-p}(1-s)$                                         & \centering $0$                                                                                      & \centering $\frac{1-p}{2-p}$                                   &               \\  
\hline
\end{tabular}
\label{scenar}
\end{table}

We first note that there are $4$ scenarios to consider: (1) both $\phi^t(\epsilon_0) \cap \Delta_1$ and $\phi^t(\epsilon_0) \cap \Delta_2$ evolve into $\epsilon_T$, (2) the former does, but the latter does not, etc. The probability distribution corresponding with scenario (1) is the same as that given by equations \ref{pm} (see equations \ref{p0}, \ref{assume}, \ref{white}). Note that each scenario's probability distribution is computed using the general equation \ref{p0} since we no longer assume perfect mixing. For scenario (2), we have $\rho_1 = 1, \rho_2 = 0$ (see equations \ref{p0}, \ref{assume}, \ref{white}, \ref{green}), and so on. The scenario in which neither evolves into $\epsilon_T$ must be dismissed since it conflicts with our initial and final boundary conditions. Our method is to compute the probability distribution for each individual scenario and average over them (weighted by each scenario's probability). The result (see Table \ref{scenar}) is:
\begin{eqnarray}
\rho_1^b &=&  \frac{1-p(1-s)}{2-p}, \nonumber\\
\rho_2^b &=&  \frac{1-p s}{2-p} \label{bp}.
\end{eqnarray}
Neither $\rho^b_1$ nor $\rho^b_2$ can reach unity or zero even in the limit $s \rightarrow 0,1$. Note from the ``denominator'' row of Table \ref{scenar} that our analysis is not defined at $s=0, 1$, hence the limit. Our claim is that (for fixed $p$) there is a fixed non-zero lower bound on $\rho^b_1$ for arbitrarily small but non-zero $s$. This bound on our certainty was not exhibited by equations \ref{pm} for the case of perfect mixing.

The entropy of $\rho^b $ is:  
\begin{equation}
  Z^b = ln(2-p)  -\frac{1-p(1-s)}{2-p} ln\left(1-p(1-s)\right) -  \frac{1-p s}{2-p} ln(1-p s). \label{zb}  
\end{equation}
 We see that $Z$ of equation \ref{pz} is the special case $p = 1$ of equation \ref{zb} for $Z^b$. Figure \ref{entPlot} shows that the minimums of $Z^b$ are always at $s \rightarrow 0,1$. In this limit equation \ref{zb} provides the following lower bound:
\begin{equation}
Z^b \ge ln(2-p) - \frac{1-p}{2-p}ln(1-p).
\end{equation}
This lower bound was zero for the special case of perfect mixing.  

Qualitatively, then, our uncertainty principle can be stated as follows. There exists a fixed bound on our certainty that the system will be in $\Delta_2$, even in the limit of experiments designed to have all phase space points in $\epsilon_0$ evolve into $\Delta_2$.

\section{The double slit experiment and the time reversed fire alarm} 
\label{ds}
Let one of the two slits of the double slit experiment be open and the other closed. If the closed slit is opened the detection frequency at some points on the detection screen will paradoxically decrease. Although generally regarded as one of the most puzzling features of quantum mechanics, we now show that this is merely a feature of a reversed thermodynamic arrow of time.

As early as 1975, Schulman \cite{schul75} showed that the thermodynamic arrow can be equivalently stated as the difference between instructions for calculating probabilities of future evolutions and those of past evolutions. Let $A^i$ be a macroscopic state comprising a region of phase space having measure $\mu (A^i)$, and let the equal a priori probability density at any point in this region be $\rho^i_a = 1/\mu(A^i)$. Given that the system is in state $A^i$ at time $s$ we are instructed to assume that $\rho^i_a$ holds when calculating the probability of the system being in a state $A^j$ at time $t >s$. This conditional probability is computed by:
\begin{equation}
P(A^j_t | A^i_s) \stackrel{(t>s)}{=} \rho^i_a \mu(A^j \cap \phi^{(t-s)} (A^i)), \label{forward}
\end{equation}   
where the subscript on a state $A$ indicates the time at which the system is found in this state, and $\phi^{(t-s)}$ is a measure preserving evolution of all phase space points in its argument from time $s$ to $t$. 

On the other hand, when calculating probabilities of evolutions into the past we are instructed {\sl not} to use $\rho^i_a$. Schulman suggests, instead, to use Bayesian inference. First, hypothesize an a priori probability distribution $q^j$ over the macroscopic states at an earlier time $t<s$ ($s$ is present time) from which the system may have evolved. Given the present state, $A^i_s$, the probability of a past (time $t$) state is then computed from: 
\begin{equation}
P(A^j_{t} | A^i_s) \stackrel{(t<s)}{=} \frac{P(A^i_s | A^j_{t}) q^j}{N}, \label{backword}
\end{equation}  
where $P(A^i_s | A^j_{t})$ can be computed using equation \ref{forward}, and $1/N$ is a normalization factor given by:
\begin{equation}
N \stackrel{(t < s)}{\equiv} \sum_k P(A^i_s | A^k_{t}) q^k.
\end{equation}

We intend to compare the above classical calculation of probabilities of past states with the calculation of probabilities of future detections in the double slit experiment. Increase $P(A^i_s | A^m_{t}) q^m$ by an amount $\Delta_m \ge 0$, $\forall m$. This corresponds in our analogy with the double slit experiment to opening the closed slit. It is clear that this may decrease the expression in equation \ref{backword} which corresponds to a detection probability in the double slit experiment.

The feature considered here need not be understood through the above equations and is, in fact, frequently experienced on a day to day basis. 

I was recently working at my desk when the fire alarm began sounding. I momentarily feared that a fire had triggered the alarm. A few seconds later, I remembered receiving an email warning us about a drill scheduled that day, and so concluded--implicitly using Bayesian inference-- that there was no fire. The conventional perspective in physics says that the fire drill merely introduced another mechanism by which the alarm could have been triggered and so could not have reduced the probability of the fire. This conventional perspective inconspicuously forbids placing boundary conditions, such as the sounding of the alarm, \emph{after} the event whose probability is to be calculated: to rule out the effect of scheduling the drill on the probability of a fire one must examine all scenarios in which the drill was and was not scheduled, \emph{including} those for which there was no alarm. In its (inconspicuous) prohibition of final boundary conditions this conventional perspective is time asymmetric, and I discard it-- at mortal risk-- in insisting on considering the sounding of the alarm as a boundary condition.

\begin{figure} 
\center
\includegraphics[scale = .42]{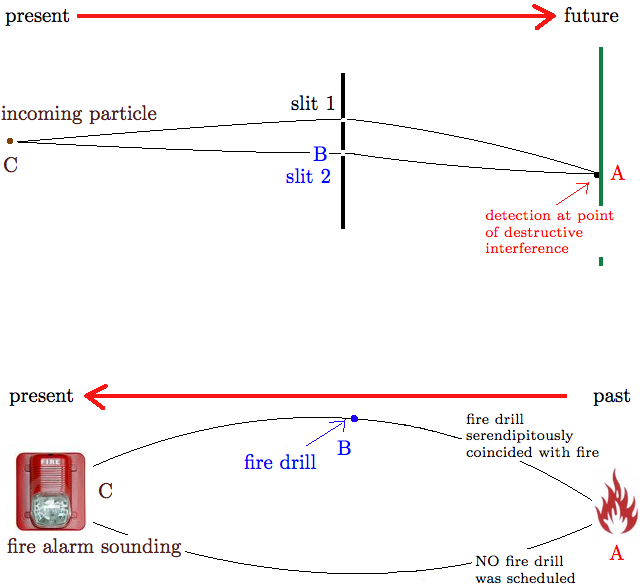}
\caption{Analogy between double slit experiment and (time reverse of) fire alarm. Events labeled by the same letter are analogous. }
\label{Analog}
\end{figure}

For a careful breakdown of the analogy, Figure \ref{Analog} recounts the fire alarm in reverse order relative to the double slit experiment. Having heard a fire alarm, there are two trajectories the universe can take into a past in which an unintended fire had started. One, in which the fire alarm is triggered by the fire, and another in which the alarm is triggered by a fire drill serendipitously scheduled on the day of the fire. These two trajectories into a past in which an unintended fire had started are analogous to the particle trajectories through the two slits into a future in which the aforementioned detection occurs. To see the analogy imagine we are uncertain of the existence of the fire. When trajectories involving a scheduled drill are ruled out, the probability of the unintended fire rises (through Bayesian inference), much as the detection probability rises when one of the slits is closed.

\section{Discussion}

Most agree that the two main themes of this paper, quantum measurement and time's arrow, are subtle topics. Attention to subtlety in physics is, to some extent, compromised by the larger emphasis some place on issues that more directly impact on theory's ability to correctly predict empirical results. 

A subtle coincidence in Newtonian theory is the similar results one obtains when one either applies an acceleration to an observer's frame of reference, or applies a uniform gravitational field to the observer's background. This subtle coincidence bore no direct impact whatsoever on Newtonian theory's ability to correctly predict empirical results. Yet, we know that consideration of this coincidence led to general relativity. In a moment we allude again to how subtle connections made between electric and magnetic fields, fluids, and light motivated the discovery that light is an electromagnetic wave. 
 
Some in science are delighted by the unfamiliar and are happy to embrace it. Feynman's attitude towards quantum mechanics is a good example. Others view science as the constant quest to understand initially unfamiliar phenomena by associating them with the more familiar. An extreme example in this group are Pythagoreans\footnote{According to legend, Pythagoreans were so convinced that all phenomena in the universe can be reduced to whole numbers and their ratios that they threw Hippasus, the discoverer of irrational numbers, overboard while out at sea. Thousands of years later Pythagoreans were somewhat vindicated: the very successful Dedekind cuts have provided a definition of irrational numbers that appeals only to (the more familiar) rational numbers.}. 

History is replete with examples of the success of this latter view. We are all aware of Newton's successful attempt to understand the unfamiliar trajectory of planets using that of apples. We are also aware that common features that electromagnetism and light share with the familiar fluids, led to the discovery that light is an electromagnetic field.

 
Feynman \cite[chap. 6]{feyn67}, \cite[chap. 3]{feyn06} claims that while we have grown accustomed to understanding foreign phenomena by associating them with the more familiar, this recipe will not work in quantum mechanics. It is largely the enigmatic feature of the double slit experiment, discussed in section \ref{ds}, that drives him to this claim. That we were able to understand this enigmatic feature in terms of familiar, albeit time reversed, scenarios of a fire alarm, contradicts Feynman's claim.

Another feature that mystified Feynman was disappearance of the interference pattern when the electron's position is measured at the slits. Some insight into this feature can be gained from section \ref{uncertainty}. Probability distributions were added non-distructively in that section only under the assumption that points in $\Delta_1$ are sufficiently displaced in phase space from points in $\Delta_2$ such that their probabilities of evolving into $\epsilon_T$ are independent. Probabilities of points within the same $\Delta$ to evolve into  $\epsilon_T$, however, are correlated. 

In other words, probability distributions are added non-distructively only when the paths are initially sufficiently displaced to expect a butterfly effect. When electron position is not measured at the slits it is not unreasonable to expect that the course of the universe is not ultimately altered by which slit the electron passes through, so long as it reaches the same place on the detection screen. Probability distributions of paths through each slit may no longer then be added non-destructively as was done is section \ref{uncertainty}. A measurement of which slit the electron passes through entails that the world macroscopically changes depending on the measurement result. A butterfly effect is then expected and the probability distributions can be added non-destructively.

\acknowledgments
I am grateful to Michael Thorn and David Miller for useful discussions and comments.

\appendix
\section{Strict Bayesianism}

\label{Caves}
In strict Bayesianism, probabilities are not defined in terms of measurement frequencies, but as degrees of belief. Caves et al. \cite{caves02} utilize a definition of probability, which they call dutch book consistency, that makes no mention of measured frequencies. They also maintain that quantum states represent states of knowledge as opposed to objective states. Accounting for the absence of superposition (indefinite) outcomes of measurement is then less problematic since the state vector is not objective. Finally, the chief challenge of Bayesianism, the search for methods to translate information into probability assignments, is fortuitously resolved by Gleason's theorem \cite{gleason57}. For these reasons, they argue, Bayesianism is especially well suited for quantum probabilities.

We, on the other hand, do not think it is necessary to abandon the concept of an objective state. We pointed out in section \ref{mp} that the absence of superposition outcomes of measurements is due to the two time boundary condition. We have yet, though, to account for Born's rule.

Consider the ensemble \{$p_i, \psi_i$\} of possible states of system and environment. Let the only assumption we make about the distribution in this ensemble be that it only include states that satisfy the two time boundary condition. Presumably then, $p_i$ will be zero for non-``special'' states. We define a density matrix \emph{based on} this ensemble. In the ensemble basis, Born's rule follows simply from the definition of the density matrix itself. Born's rule need only be accounted for for measurements in other bases. But the two time boundary condition leads us to believe that the ensemble basis is always the measurement basis itself (that states always happen to evolve into one of the basis vectors of ensuing measurements). And so Born's rule is accounted for.

The weakness in our argument is that the above density matrix, which incorporates restriction to special states, is not the conventionally used density matrix. We view the latter as a course grain representation of the former. The weakness then is our assumption that while the final boundary condition obviously influences evolution of the ensemble, it does not influence evolution of the density matrix derived from this ensemble: that the course grained and primitive density matrices undergo the same evolution. This weakness is currently being addressed in an independent research effort.

\end{document}